\documentclass[]{article}
\usepackage[english]{babel}
\usepackage{graphicx,enumerate}
\usepackage[tbtags]{amsmath}

\begin{document}
\title{Minimum energy required to copy one bit of information}
\author{Marcin Ostrowski\footnote{e-mail: ostrowsk@ics.p.lodz.pl}}
\maketitle

\begin{abstract}
In this paper, we calculate energy required to copy one bit of useful information in the presence of thermal noise.
For this purpose, we consider a quantum system capable of storing one bit of classical information, which is initially
in a mixed state corresponding to temperature $T$. We calculate how many of these systems must be used to store useful
information and control bits protecting the content against transmission errors. Finally, we analyze how adding these
extra bits changes the total energy consumed during the copying.
\end{abstract}

\section{Introduction - Landauer principle}
What is the relationship between information and heat (energy)? Does a machine that performs calculations (such as
a computer) have to produce heat? These questions are purely fundamental but they are also important from a technical point
of view. Firstly, as computer users, we want to pay less for the electricity we consume. Secondly, the problem of
cooling modern electronic systems with high packing density of components is a great challenge for engineers. 

\noindent
The relationship between heat $Q$ (and the thermodynamic entropy $S$) and information had occupied the attention of researchers for
a long time before the invention of the computer. Let us recall here the problem of Maxwell's demon separating  faster
particles from the slower which resulted in the flow of heat from a colder to a hotter reservoir. These considerations
led to the suggestion that the measure (gain of information) always requires the increase of thermodynamic entropy [1], [2]. 

\noindent
According to the present state of knowledge, information processing (calculations) may take place without the production
of heat. In this case, we are talking about reversible computations. The main feature of reversible computations is the ability
to recover input data based on output data. The machine fully capable of implementing this type of task would be a quantum
computer [3], [4].

\noindent
However, the modern digital machine is not reversible. Moreover, apart from the calculations, it must perform other functions
(associated with the calculations) as well as reading of data, copying of data, etc.. In 1961, Rolf Landauer showed in his
work [5] that any logically irreversible manipulation of information, such as erasing the information, must result in an
increase in entropy of the apparatus or the environment. Erasing one bit of information should result in the production
of heat at the order of $kT\ln2$. This is often called the ''Landauer principle''.

\noindent
The present paper relates to those considerations, but it does not recalculate the Landauer's result using a different method.
In our work, we consider another (though related) issue. We wonder how much energy should be used to make a copy of one bit useful
information when we do not have a pure medium on which we could save a copy, that is, if there is a necessity, to overwrite
the original thermal noise appearing on the medium. Therefore, we assume that information is protected against errors due to
thermal noise by the presence of additional control symbols (redundancy). We investigate whether it is better to use
a small number of control symbols with a greater energy expenditure per bit (higher signal to noise ratio) or the other way round
(ie. use a large number of control symbols at low signal to noise ratio). For this purpose, we introduce the concept of
quantum copier that makes copies of information between subsystems. The main result of the work (presented in the last section)
is limitation for the minimum energy required to copy one bit similar to Landauer's one.

\section{Quantum copier}

Consider a physical (quantum) system that stores one bit of classical information. Let logic state ''0'' correspond to
the quantum state $\vert0\rangle$ and logic state ''1'' correspond to the quantum state $\vert1\rangle$.
States $\vert0\rangle$ and $\vert1\rangle$ are orthogonal to each other (which ensures their distinguishability in the
measurement), normalized and form the basis of a quantum system.

\noindent
Let us consider two such systems which will be called $A$ and $B$, respectively. Next, we construct a quantum operation that
copies the states ''0'' and ''1'' from the system $A$ to system $B$ according to the following scheme:
\begin{eqnarray}
\vert0\rangle_A\vert\textrm{pm}\rangle_B\rightarrow\vert0\rangle_A\vert0\rangle_B\\
\vert1\rangle_A\vert\textrm{pm}\rangle_B\rightarrow\vert1\rangle_A\vert1\rangle_B
\end{eqnarray}
where the state $\vert\textrm{pm}\rangle_B$ means a given state of the system $B$. As we shall see, copy could be done
properly only if the system $B$ is initially in a particular state, which here is denoted by the abbreviation ''pm'', which stands for
''pure medium''.

\noindent
The choice of states $\vert0\rangle$ and $\vert1\rangle$ as basis vectors (and not as some combination
of the basis states) is only a simplification of the notation does not restrict the generality of our considerations.

\noindent
In order to fully determine the unitary operation acting on subsystems $A$ and $B$ we have to introduce vector $\vert\textrm{um}\rangle_B$
(from the words ''unprepared medium'') orthogonal to $\vert\textrm{pm}\rangle_B$ and determine the action of the operation on it.
If we did it as follows\footnote{
In the further part of the work, we will use shorter notation of states (i.e. without lower indices $A$, and $B$). States
$\vert0\rangle_A\vert1\rangle_B$ will be denoted by $\vert0\rangle\vert1\rangle$ or even $\vert0\,1\rangle$.}:
\begin{eqnarray*}
\vert0\rangle\vert\textrm{um}\rangle\rightarrow\vert0\rangle\vert0\rangle\\
\vert1\rangle\vert\textrm{um}\rangle\rightarrow\vert1\rangle\vert1\rangle
\end{eqnarray*}
then our operation would be able to copy the initial state of subsystem $A$ to subsystem $B$, regardless of the initial
state of medium $B$. Unfortunately, this type of operation is not unitary. This would be an irreversible process erasing
information about the initial state of $B$. Thus, we have to design our system as follows:
\begin{eqnarray}
\vert0\rangle\vert\textrm{um}\rangle\rightarrow\vert0\rangle\vert1\rangle\\
\vert1\rangle\vert\textrm{um}\rangle\rightarrow\vert1\rangle\vert0\rangle
\end{eqnarray}
This means that for copying operation (which will be called the quantum copier) to work correctly, the second entry must
always be in the state $\vert\textrm{pm}\rangle$. Otherwise errors will occur.

\noindent
Equations (1)-(4) fully define the operation of the copier as a unitary operation acting over $H_A\otimes H_B$.
Let us emphasize that the copier can copy only the base states ($\vert0\rangle_A$ and $\vert1\rangle_A$), not their superpositions.
Copying of any state is impossible by the no-cloning theorem.

\section{Copy to the unprepared medium (in the presence of noise)}

Let us examine what the result of copying will be when the medium $B$ is initially unprepared. For this purpose, we choose the
initial state of $B$ as a mixed state in the form:
\begin{equation}
\rho_B=(1-b)\vert\textrm{pm}\rangle\langle\textrm{pm}\vert+b\vert\textrm{um}\rangle\langle\textrm{um}\vert
\end{equation}
where $b$ can be identified with a probability of incorrect preparation of the medium (and therefore erroneous copying).

\noindent
In the case of $A$, we assume that with probability $p_0$ the system stores a ''0'', and with probability of $p_1$ it stores
a  ''1''. Therefore, we assign the system $A$ a mixed state in the following form:
\begin{eqnarray}
\rho_A=p_0\vert0\rangle\langle0\vert+p_1\vert1\rangle\langle1\vert
\end{eqnarray}

\noindent
Subsystems $A$ and $B$ are together in a state which can be written as: 
\begin{multline}
\rho_{AB}=\rho_A\otimes\rho_B=p_0(1-b)\vert0\rangle\vert\textrm{pm}\rangle\langle\textrm{pm}\vert\langle0\vert+
p_0b\vert0\rangle\vert\textrm{um}\rangle\langle\textrm{um}\vert\langle0\vert+\\+
p_1(1-b)\vert1\rangle\vert\textrm{pm}\rangle\langle\textrm{pm}\vert\langle1\vert+
p_1b\vert1\rangle\textrm{um}\rangle\langle\textrm{um}\vert\langle1\vert
\end{multline}

\noindent
After copying information (after applying the unitary operation $U_c$ described by the equations (1)-(4)) we obtain:
\begin{multline}
\rho_{AB}'=U_c\rho_{AB}U_c^+=p_0\vert0\rangle\langle0\vert_A[(1-b)\vert0\rangle\langle0\vert_B+b\vert1\rangle\langle1\vert_B]
+\\+
p_1\vert1\rangle\langle1\vert_A[(1-b)\vert1\rangle\langle1\vert_B+b\vert0\rangle\langle0\vert_B]
\end{multline}
and after counting the partial trace:
\begin{eqnarray}
\rho_A'&=&p_0\vert0\rangle\langle0\vert+p_1\vert1\rangle\langle1\vert\\
\rho_B'&=&[p_0(1-b)+p_1b]\vert0\rangle\langle0\vert+[p_0b+p_1(1-b)]\vert1\rangle\langle1\vert
\end{eqnarray}

\noindent
Let us look at specific cases:
\begin{eqnarray}
\rho_B'&=&p_0\vert0\rangle\langle0\vert+p_1\vert1\rangle\langle1\vert\quad\textrm{for}\quad b=0\\
\rho_B'&=&\tfrac{1}{2}\vert0\rangle\langle0\vert+\tfrac{1}{2}\vert1\rangle\langle1\vert\quad\textrm{for}\quad b=\tfrac{1}{2}\\
\rho_B'&=&p_1\vert0\rangle\langle0\vert+p_0\vert1\rangle\langle1\vert\quad\textrm{for}\quad b=1
\end{eqnarray}

\noindent
Thus, we have obtained what we expected. For $b=0$ we obtain a faithful copy, for $b=\tfrac{1}{2}$ we obtain a complete lack of
information copied from $A$ to $B$, for $b=1$ we get a faithful copy with negation (the states $\vert\textrm{pm}\rangle_B$
and $\vert\textrm{um}\rangle_B$ turn its roles).

\section{Relationship with Shannon's binary information channel}

We will treat states of $A$ (''0'' and ''1'') as the realization of a stochastic process $X$ and states of $B$ (after
copying) as the  realization of stochastic process $Y$. The relationship between these two processes can be depicted by the
graph in Figure~1.

\begin{figure}[h]
\begin{center}
\includegraphics[width=4.5cm]{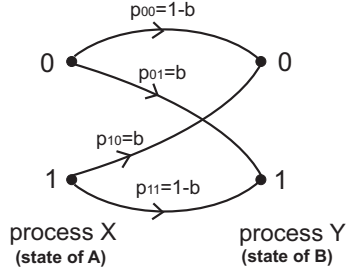}

\end{center}
\caption{Relation between the processes $X$ and $Y$. The $p_{00}$ and $p_{11}$ - probability of correct transmission,
$p_{01}$, $p_{10}$ - probability of erroneous transmission}
\end{figure}

\noindent
It is a complete analogy with the Shannon's, binary, noisy information channel. The process $X$ correspond to the process
at the entrance to the channel, the process $Y$ in the output.

\noindent
In the next step, we calculate transinformtion $I(X,Y)$ for these processes (i.e. the amount of knowledge about $X$ which
is contained in $Y$):
\begin{equation}
\label{eq_shannon}
I(X,Y)=\sum_{i,j}P(X_i,Y_j)\log_2\frac{P(X_i,Y_j)}{P(X_i)P(Y_j)}
\end{equation}
where the probabilities for the process $X$ are given by:
\begin{equation*}
P(X_0)=p_0 \quad P(X_1)=p_1
\end{equation*}
while for the process $Y$ are in the form:
\begin{equation*}
P(Y_0)=p_0(1-b)+p_1b \quad P(Y_1)=p_0b+p_1(1-b)
\end{equation*}
while the total probabilities:
\begin{eqnarray*}
P(X_0,Y_0)=p_0(1-b) \quad\quad P(X_0,Y_1)=p_0b\\
P(X_1,Y_0)=p_1b \quad\quad P(X_1,Y_1)=p_1(1-b)
\end{eqnarray*}

\noindent
Let us calculate transinformation (\ref{eq_shannon}) for the case $p_0=p_1=\tfrac{1}{2}$, since then in $A$, we have
exactly the 1bit of information. Then:
\begin{equation}
\label{eq_inofs}
I(X,Y)=1+(1-b)\log_2(1-b)+b\log_2b.
\end{equation}

\section{Medium as a system with two states of energy}

For simplicity, we assume that $\vert\textrm{pm}\rangle_B=\vert0\rangle_B$ and $\vert\textrm{um}\rangle_B=\vert1\rangle_B$.
In addition, we assume that they are stationary states of the free hamiltonian of $B$. Otherwise, these states would evolve
in time and reading of information (stored in them) at any moment of time would be difficult. Thus, let $E_0$ and $E_1$ denote
energy of state $\vert0\rangle_B$ and energy of the $\vert1\rangle_B$, respectively.

\noindent
Our main assumption is that before copying the system $B$ was in contact with the thermostat at a temperature of $T$
long enough to reach thermodynamic equilibrium. In practice, this is a natural assumption, since any real device works at a temperature
different from the absolute zero.\footnote{In practice, some systems may be in a quasiequilibrium state, while a state of full equilibrium
is possible to achieve only after a very long time. However, this situation will not be considered here.}

\noindent
Probability of filling the levels $\vert0\rangle$ and $\vert1\rangle$ at temperature $T$ are respectively given by:
\begin{eqnarray}
P_0&=&\frac{1}{1+\exp(-\beta\Delta)},\\
P_1&=&\frac{\exp(-\beta\Delta)}{1+\exp(-\beta\Delta)}\,=\,b
\end{eqnarray}
where $\Delta=E_1-E_0$.

\noindent
It is worth noting that $b=P_1$ only for $b\in\langle0;\tfrac{1}{2}\rangle$. For $\beta\Delta>>1$ error rate $b$ is very low,
while for $\beta\Delta\rightarrow0$ error rate $b$ tends to $\tfrac{1}{2}$. The most important conclusion is that medium $B$ in thermal
equilibrium is useful for copying if the energies of the levels differ (ie. $\Delta\neq0$). Otherwise, $P_0=P_1=1/2$ and (under
considerations presented in Section 3) the medium is totally useless.

\section{Redundancy}

In order to copy useful information, when $b\neq 0$, we have to protect transmission against possible errors.
To this end, we must encode useful information by adding a certain amount of control symbols. Let us denote the number of all
symbols by $n$ (one symbol carrying one bit of useful information and $n-1$ symbols are control symbols). In practice,
we should have not only one system $A$ but the whole series of systems in an amount of $n$. We also need $n$ systems $B$ in the
states corresponding to a temperature of $T$, where we finally put ''dirty'' copies from systems $A$.

\noindent
According to the Shannon theorem [6], we can conclude that for transfer of one bit of useful information, $n=I(X,Y)^{-1}$ bits
are necessary, where $I(X,Y)$ is given by formula (\ref{eq_inofs}). We do not examine here any specific coding procedure
(the values that control bits must take are not defined).

\section{Calculations, graphs, conclusions}

Let us calculate the energy that must be used to copy $n$ symbols from systems $A$ to systems $B$. It is calculated as follows:
\begin{equation}
\label{eq_q1b}
W=n(\tilde{E}'-\tilde{E})
\end{equation}
where $\tilde{E}=\Delta P_1=\Delta b$ is the average energy of $B$ before copying (i.e., at a temperature of $T$) calculated
for $E_0=0$ and $b\in\langle0;\tfrac{1}{2}\rangle$. Energy $\tilde{E}'=\tfrac{1}{2}\Delta$ is the average energy of $B$ after copying.
We assume here that the probabilities of occurrence of zeros and ones in the $B$ are equal and amount to 1/2.\footnote{
This follows from the following facts: $p_0=p_1=\tfrac{1}{2}$. In addition, we assume that among the control 
symbols added before copying is the same number of ''0'' and ''1''. This type of code are usually the most optimal.
During copying from $A$ to $B$ probability of distortion from ''0'' to ''1'' and from ''1'' to ''0'' are the same (Fig.~1).
This results in conservation of symmetry between the zeros and ones on the $B$.}
This gives the energy (\ref{eq_q1b}) equal to:
\begin{equation}
\label{eq_q2b}
W(\beta,\Delta)=\frac{\Delta(\tfrac{1}{2}-b)}{1+(1-b)\log_2(1-b)+b\log_2b},
\end{equation}
where $b$ is given by Eq.~(17).

\noindent
Energy (\ref{eq_q2b}) is an increasing function of $\Delta$. This means that it is less expensive to encode the information
at low signal to noise ratio with a large amount of control symbols $n$. In the borderline case, we obtain:
\begin{equation}
W_{min}=\lim_{\Delta\rightarrow0}W(\beta,\Delta)=\frac{\log4}{\beta}
\end{equation}

\begin{figure}[h]
\begin{center}
\includegraphics[width=5cm]{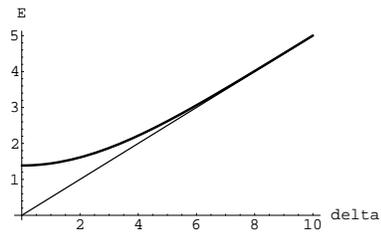}

\end{center}
\caption{Energy used for copying one bit of useful information as function of $\Delta$ for $\beta=1$. The straight line
(added for comparison) corresponds to the value $\Delta/2$.}
\end{figure}

\noindent
Energy (\ref{eq_q2b}) does not correspond to the scattered heat (as in Landauer). Our copy machine works reversibly.
The copy obtained in the $B$ still contains information about the state of medium $B$ before copying. This information can be
recovered by comparing the "dirty" copy with original one saved on $A$. Thus, there is no erasure of information. Erasure may occur
in the next phase - on machine decoding and removing copying errors.

\section*{References}
[1] L.~Szilard, Z~f~Physik, {\bf 53}, 840, (1929),

\noindent
[2] C.~H.~Bennett, Int. J. Theor. Phys., {\bf 21}, 905, (1982),

\noindent
[3] Feynman R. P. {\it Simulating physics with computers}, Int. J. Theor. Phys. 21:6/7.pp. 467-488 (1982),

\noindent
[4] Feynman {\it Feynman lectures on computation}, (1996),

\noindent
[5] R.~Landauer, IBM J. res. Develop., {\bf 5}, 183, (1961),

\noindent
[6] Shanonn C. E. {\it The Mathematical Theory of Communication}, The Bell System Technical Journal, {\bf 27}, 1948

\end{document}